\documentclass[runningheads]{llncs}
\usepackage[T1]{fontenc}
\usepackage{multirow}
\usepackage{array}
\usepackage{amsmath}
\usepackage{amssymb}
\usepackage{url}
\usepackage{graphicx}
\begin{document}
\title{Decoupling Predictions in Distributed Learning for Multi-Center Left Atrial MRI Segmentation}
\titlerunning{Decouple Predictions in Distributed Learning for Multi-Center LA MRI}
%
\author{Zheyao Gao\inst{1} \and
Lei Li\inst{3} \and Fuping Wu\inst{1,2} \and Sihan Wang\inst{1} \and Xiahai Zhuang\inst{1}\thanks{Xiahai Zhuang is corresponding author. This work was funded by the National Natural Science Foundation of China (grant no. 61971142, 62111530195 and 62011540404)
and the development fund for Shanghai talents (no. 2020015)}}

\institute{School of Data Science, Fudan University, Shanghai, China \and Department of Statistics, Fudan University, Shanghai, China \and Department of Engineering Science, University of Oxford, Oxford, UK \\
\url{www.sdspeople.fudan.edu.cn/zhuangxiahai/}}
\authorrunning{Z. Gao et al.}
\maketitle              
\begin{abstract}
Distributed learning has shown great potential in medical image analysis. It allows to use multi-center training data with privacy protection. 
However, data distributions in local centers can vary from each other due to different imaging vendors, and annotation protocols. 
Such variation degrades the performance of learning-based methods. 
To mitigate the influence, two groups of methods have been proposed for different aims, i.e., the global methods and the personalized methods.
The former are aimed to improve the performance of a single global model for all test data from unseen centers (known as generic data); while the latter target multiple models for each center (denoted as local data). 
However, little has been researched to achieve both goals simultaneously. 
In this work, we propose a new framework of distributed learning that bridges the gap between two groups, and improves the performance for both generic and local data. Specifically, our method decouples the predictions for generic data and local data, via distribution-conditioned adaptation matrices. 
Results on multi-center left atrial (LA) MRI segmentation showed that our method demonstrated superior performance over existing methods on both generic and local data. Our code is available at \url{https://github.com/key1589745/decouple_predict}

\end{abstract}
\section{Introduction}

Left atrial (LA) segmentation from MRI is essential in the diagnosis and treatment planning of patients suffering from atrial fibrillation, but the automated methods remain challenging~\cite{LAseg}. 
Deep learning has demonstrated great potential, provided a large-scale training set from multiple centers, which nevertheless impedes the advance due to the concern of data privacy. 
Distributed learning has then gained great attention as it trains a model on distributed datasets without the exchange of privacy-sensitive data between centers~\cite{FedAvg}.
Recently, swarm learning~\cite{swarm}, a decentralized distributed learning method, has been proposed for medical image analysis. This new model follows the training paradigm of  federated learning  (FedAvg) but without a central server. The communication between centers is secured through the blockchain network~\cite{blockchain}, which further guarantees fairness in distributed learning.

\begin{figure}[t]
  \centering 
  \begin{tabular}{ccccc}
    \includegraphics[width=0.18\linewidth,height=0.18\linewidth]{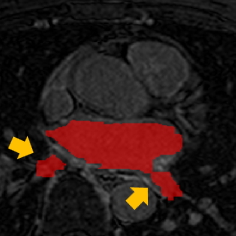}&
    \includegraphics[width=0.18\linewidth,height=0.18\linewidth]{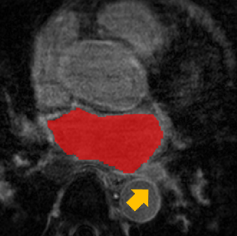}&
    \includegraphics[width=0.18\linewidth,height=0.18\linewidth]{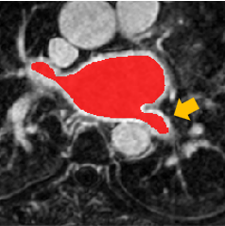}&
    \includegraphics[width=0.18\linewidth,height=0.18\linewidth]{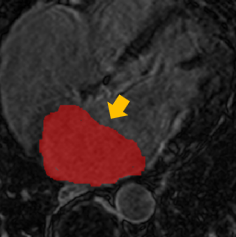}&
    \includegraphics[width=0.18\linewidth,height=0.18\linewidth]{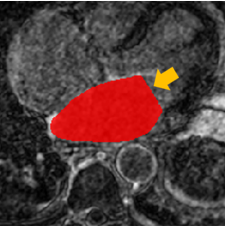}\\(a)&(b)&(c)&(d)&(e)\end{tabular}\\[-2ex] 
  \caption{Illustration of label skew in different centers. Examples are selected from public datasets~\cite{data1,data2}. Yellow arrows point out the segmentation bias: {(a-c)} show different protocols for segmenting pulmonary veins, {(d-e)} demonstrate inconsistency for the LA cavity.}
\label{fig:intro}
\end{figure}

Although the above methods have provided solutions for the problem of  privacy and fairness, the performance could be degraded due to the non-independently identical distribution (IID)-ness of medical data~\cite{FedAvg,noniid}. For left atrial (LA) MRI segmentation, the heterogeneous of data distribution could be particularly worse in two aspects. One is the divergence of image distributions (also known as feature skew), originated from the difference in imaging protocols, the strength of magnetic field, or demography. The other is the inconsistency of manual labeling (referred to as label skew) due to different annotation protocols or rater variations, as manual segmentation of LA MRIs is difficult even for experienced experts~\cite{label}. 
Fig.~\ref{fig:intro} illustrates this inconsistency from different centers.

For distributed learning under data heterogeneity, solutions could be categorized as global methods or personalized methods. Global methods~\cite{FedProx,FedCurv,scaffold} usually regularize the local training to avoid the divergence of model updating in each center. Personalized methods~\cite{FedBN,FedRep,ditto} propose to keep distribution related parameters local or develop two sets of models through distillation. In general, global methods can perform well on generic data~\cite{noniid} while personalized methods have shown better results on local data.
A recent work~\cite{bridge} has been proposed to approach the two seemingly contradictory objectives simultaneously with two predictors for image classification.
Different from \cite{bridge}, we developed a novel distributed learning framework for LA segmentation, based on the variational Bayesian framework, which bridges the objectives of global methods and personalized methods.

In the segmentation task with non-IID data, little research has studied the influence of label skew caused by segmentation bias and variation in different centers.
Therefore, we resort to the remedies for noisy label learning by considering the segmentation bias in each center as noises in the annotation. A large number of works~\cite{MAE1,MAE,lq} have studied the conditions for robust loss functions and proposed simple yet effective solutions for learning with label noise.
Several studies~\cite{adapt1,adapt2} have proposed to add an adaptation layer on top of the network. A recent work~\cite{cms} disentangled annotation errors from the segmentation label using an image dependent adaptation network. However, it requires annotations from multiple observers for a single case and the adaptation network could not learn the segmentation bias from labels. For our problem, we consider adapting the prediction based on the joint distribution of both image and label in each center through generative models. 

In this work, our distributed learning method, tackling both problems of feature and label skew, improves the performance for both generic and local data. 
More specifically, our method decouples the local and global predictions through adaptation matrices conditioned on joint data distribution, 
thus prevents the divergence of model in local training and adapts the prediction to be consistent with local distribution during testing. 

Our contribution has three folds. First, we propose a new distributed learning framework for LA MRI segmentation with non-IID data, which could improve the results on data from both unseen centers and centers involved in the training.
Second, we propose a distribution-conditioned adaptation network to decouple the prediction with labels during training and adapt the prediction according to local distribution in testing.
Third, We evaluate the proposed framework on multi-center LA MRI data. The results show that our method outperforms existing methods on both generic data and local data.

\begin{figure}[h]
  \centering
\begin{tabular}{cc}
    \includegraphics[width=0.34\linewidth]{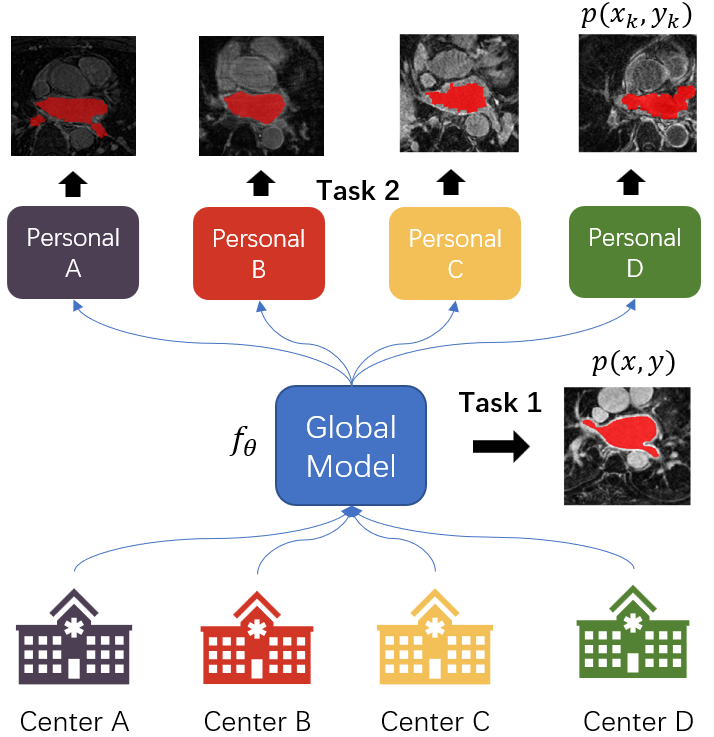}&
	\includegraphics[width=0.64\linewidth]{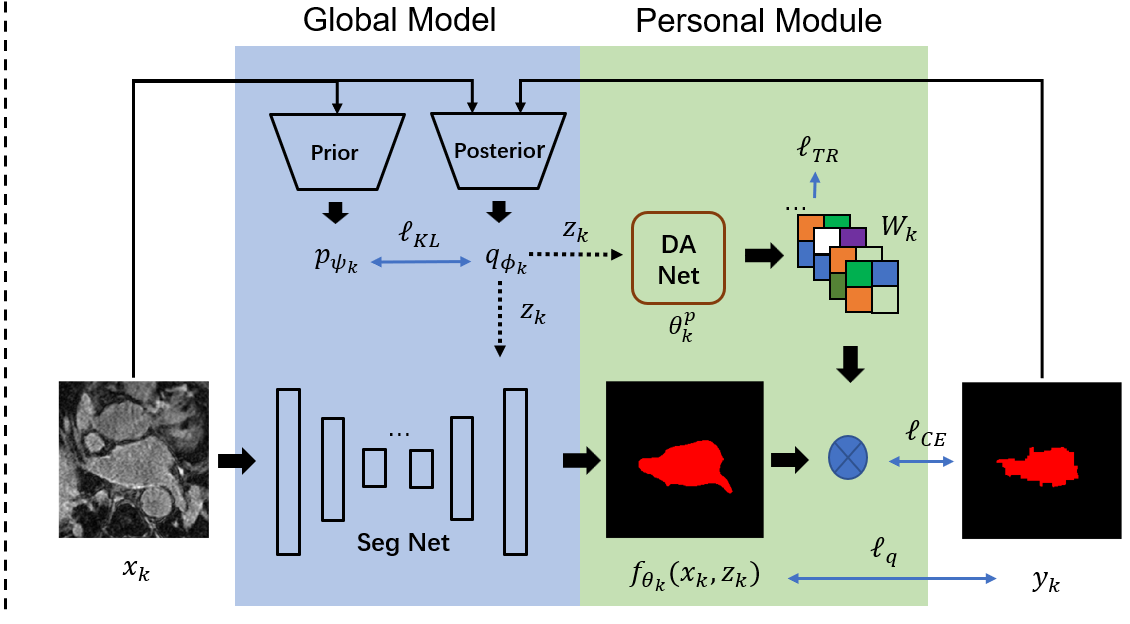}	\\
	(a)  & (b)\end{tabular}\\[-2ex]
  \caption{Illustration of the objective (a) and framework of local training (b).}
    \label{fig2}
\end{figure}

\section{Method}
Fig.~\ref{fig2} (a) illustrates two folds of our aim. One is to train a global model $f_{\theta}$  from multi-center non-IID data for Task1, and the other is to seek a personalized one for each center for Task2.
The global model is expected to achieve consistent segmentation for generic data distribution~\cite{noniid} $p(x,y)$, \textit{i.e.,} images from unseen centers;
the personalized models, by contrast, modify predictions from the global model to favor the local distribution, e.g., $p(x_k,y_k)$ of the $k$-th center. 
The proposed method achieves the two goals simultaneously.

Based on swarm learning~\cite{swarm}, our method involves periodically global aggregation and local training. In \emph{global aggregation}, each center $k$ collects locally updated parameters $\{(\theta_k)\}_{k=1}^K$ of the global model from other centers, and performs model aggregation weighted by the training size of each center ($n_k$), \textit{i.e.}, $\theta=\sum_{k=1}^K{\frac{n_k}{N}\theta_k}$ where $N=\sum_{k=1}^K{n_k}$. 
Note that the personalized modules, denoted as $\{\theta^p_k\}$, are kept locally in this stage. 
In \emph{local training}, each center then trains their own models, with the global model initialized as the aggregation result $\theta$.  

Fig.~\ref{fig2} (b) shows the proposed framework of local training, consisting of four sub tasks respectively for segmentation network ($\theta_k^s$), prior encoder ($\psi_k$), posterior encoder ($\phi_k$) and the distribution adaptation (DA) network ($\theta^p_k$). 
The former three maintain the information for updating the global model. The segmentation network generates predictions for generic data.  
Prior and posterior encoders are employed to model the joint data distribution $p(x_k,y_k)$ through a latent representation $z_k$. Thus, the DA network could output adaptation matrices $W_k$ based on local distribution.
The prediction for local data is obtained through the multiplication of adaptation matrices $W_k$ and the prediction for generic data. In the following sections, we elaborate on the derivation and training details for each part of the model.

\subsection{Modeling Joint Data Distribution}
Since local training is driven by the segmentation risk on local data distribution, $\theta_k$ would diverge from each other if the data distributions are non-IID. This will deviate the aggregated parameter $\theta$ at convergence from the optimal solution for the generic test data from Task1~\cite{noniid}. 
To mitigate this problem, we propose to modify the prediction of the global model based on local joint distribution $p_k(x_k,y_k)$. However, the label distribution is not available during testing. Therefore, we first model the latent representation $z_k$ that contains the information about joint distribution following the formulation of conditional VAE~\cite{cVAE}.

Concretely, the objective is to estimate the posterior distribution $p(z_k|x_k,y_k)$. The key idea is to find a Gaussian distribution $q_{\phi_k}(z_k|x_k,y_k)$ to approximate the true posterior by Kullback-Leibler (KL) divergence. According to~\cite{cVAE}, the lower bound (LB) of the KL divergence is derived as,
\begin{equation}
    LB= \mathbb{E}_{q_{\phi_k}(z_k|x_k,y_k)}[\log{p(y_k|x_k,z_k)}]-KL[q_{\phi_k}(z_k|x_k,y_k)||p_{\psi_k}(z_k|x_k)], \label{eq:LB}
\end{equation}
where the approximated prior $p_{\psi_k}(z_k|x_k)$ and posterior $q_{\phi_k}(z_k|x_k,y_k)$ are modeled by,
\begin{equation}
p_{\psi_k}(z_k|x_k)=\mathcal{N}\big(\mu_{prior}(x_k;\psi_k),\texttt{diag}(\sigma_{prior}(x_k;\psi_k))\big),
\end{equation}
\begin{equation}
q_{\phi_k}(z_k|x_k,y_k)=\mathcal{N}\big(\mu_{post}(x_k,y_k;\phi_k),\texttt{diag}(\sigma_{post}(x_k,y_k;\phi_k))\big).
\end{equation}
The parameters of the Gaussian distributions, $(\mu_{prior},\sigma_{prior})$ and $(\mu_{post},\sigma_{post})$ are generated by the prior and posterior network, respectively. 

The posterior distribution of the latent representation $z_k$ is approximated through maximizing Eq.~(\ref{eq:LB}). The KL term in the lower bound could be derived and maximized explicitly. The maximization of the first term is resolved through the training of segmentation network and DA network, which will be described in the following.

\subsection{Decoupling Global and Local Predictions}
To disentangle the effect of distribution shift on the prediction and avoid the parameters from drifting away during local training, we decouple the predictions for generic data with labels from local distribution. Formally, we decompose the first term in Eq.~(\ref{eq:LB}) as following,
\begin{flalign}
    \mathbb{E}_{q_{\phi_k}}[\log{p(y_k|z_k,x_k)}] =
    \mathbb{E}_{q_{\phi_k}}[\log\sum_{y\in\Omega}{p_{\theta^p_k}(y_k|y,z_k)p_{\theta^s_k}(y|x_k,z_k)}],
\end{flalign}
where $p_{\theta^p_k}(y_k|y,z_k)$ is modeled by the personalized module; $p_{\theta^s_k}(y|x_k,z_k)$ is modeled by the segmentation network; $\Omega$ is the set of all possible segmentation. Here, $p_{\theta^p_k}(y_k|y,z_k)$ is independent on $x_k$, as we assume that the difference between $y$ and $y_k$ is resulted by the shift of joint data distribution instead of $x_k$. In this way, the predictions for generic data and local data are decoupled. The DA network bridges the gap caused by the distribution shift.

To jointly train the segmentation network and DA network, we first apply cross-entropy loss $\ell_{CE}$~\cite{CE} to the multiplication of outputs from the two networks with $y_k$, \textit{i.e.}, $\ell_{CE}(W_{k}\odot f_{\theta_k}(x_k,z_k),y_k;\theta_k)$,
where $W_k\in\mathbb{R}^{C\times C\times M}$ are the adaptation matrices for all pixels. $C$ and $M$ represent the number of classes and pixels respectively. "$\odot$" here denotes the pixel-wise matrix multiplication. This loss ensures that the prediction for local data $W_k\odot f_{\theta_k}(x_k,z_k)$ is congruent local labels.

However, $\ell_{CE}$ alone is not able to separate the effect of distribution shift from the global prediction. There are many combinations of $\theta^p_k$ and $\theta^s_k$ such that given an input image, the prediction for local data perfectly matches the labels. To avoid the analogous issues, we further introduce a regularization term for the adaptation matrices and a segmentation loss that directly compares the prediction of $f_\theta$ and the label $y_k$,
\begin{equation}
    \ell_{TR}(W_{k};\theta^p_k)=\frac{1}{M}\sum_{i=1}^M{tr(W_k^i)}, \label{eq:TR}
\end{equation}
\begin{equation}
    \ell_{NR}(f_{\theta_k}(x_k,z_k),y_k;\theta_k)=\sum_{i=1}^M\sum_{j=1}^C y_k^{[i,j]}(\frac{1-f_{\theta_k}^{[i,j]}(x_k,z_k)^q}{q}), \label{eq:lq}
\end{equation}
where $W_k^i$ represent the adaptation matrix for pixel $i$, and $\ell_{TR}$ is to minimize the diagonal elements of $W_k^i$ for each pixel. The non-diagonal elements which denote label flipping probabilities become dominant. Thus, it forces the adaptation matrices to modify the prediction as much as possible. 
In Eq.~(\ref{eq:lq}), $y_k^{[i,j]}$ and $f_{\theta_k}^{[i,j]}$ denote the $j$th channel of pixel $i$ in the one-hot label and the prediction, respectively. $\ell_{NR}$ is the noise robust loss, which is similar to the generalized cross-entropy loss introduced in~\cite{lq}. $q$ is the hyperparameter to balance between noise tolerance and convergence rate. As $q\to 0$, it is equivalent to CE loss which emphasizes more on uncertain predictions, and it degrades to MAE loss~\cite{MAE}, which equally penalizes on each pixel, as $q$ approaches 1. 

$\ell_{NR}$ guarantees that the prediction of the segmentation network $f_{\theta_k}(x_k,z_k)$ is close to $y_k$, while $\ell_{TR}$ forces the adaptation matrices to modify the uncertain predictions. As the parameters of global model are obtained in a distributed training manner, the uncertain predictions are likely to be caused by the distribution shift in local data. Therefore, the effect of distribution shift is disentangled through the combination of $\ell_{NR}$ and $\ell_{NR}$.

\subsection{Training and Testing Details}
\subsubsection{Training} The local training procedure is similar to that  of probabilistic U-Net~\cite{probU}. Given the input image $x_k$ and segmentation ground truth $y_k$, the low-dimension representation $z_k\in\mathbb{R}^D$ is sampled from the estimated posterior $q_{\phi_k}$. It is broadcast to the same size of the image as a $D$-channel feature to concatenate with the last feature map of the segmentation network. In our work, it is also taken as the input of DA network to yield adaptation matrices $W_k$. To initialize the adaptation matrices and distribution encoders, we first perform warm-up training in the first few local training epochs with, 
\begin{equation}
    \ell_{warm} = \ell_{CE}(f_{\theta_k}(x_k,z_k),y_k;\theta_k)-\beta \ell_{KL}-\ell_{TR},
    \label{Eq:warm}
\end{equation}
where $\ell_{KL}$ is derived from the KL term in Eq.~(\ref{eq:LB}), weighted by the hyperparameter $\beta$. $\ell_{CE}$ and $\ell_{KL}$ in Eq.~(\ref{Eq:warm}) are to train the segmentation network and distribution encoders, respectively, such that the DA network could acquire the latent representation of the joint data distribution in the next stage. $\ell_{TR}$ initializes the adaptation matrices to be diagonal dominant such that it does not modify the prediction in the beginning. The overall loss applied during the main training procedure is derived as,
\begin{equation}
    \ell = \ell_{CE} + \ell_{NR} +\alpha \ell_{TR} - \beta \ell_{KL},
\end{equation}
where $\alpha$ is applied to balance the regularization for the adaptation matrices. 
\subsubsection{Testing} During testing, the latent representation $z$ is randomly sampled from the prior $p_{\theta_k}$. The prior distribution could resemble   the posterior distribution in a way that reflects different modes of the segmentation results in each center due to the the KL loss~\cite{probU}. Thus, it could also be applied to model the effect of distribution shift on the prediction.




\section{Experiment}
\subsection{Dataset and Preprocessing}

We evaluated our method using LA segmentation on late gadolinium enhanced (LGE) MRIs, collected from three centers, \textit{i.e.} Beth Israel Deaconess Medical Center (Center A), Imaging Sciences at King’s College London (Center B) and Utah School of Medicine (Center C, D). The datasets were obtained from MICCAI 2018 Atrial Segmentation Challenge~\cite{data1} and ISBI 2012 Left Atrium Fibrosis and Scar Segmentation Challenge~\cite{data2}. 

For Center A and B, we selected 15 cases for training and 5 cases as local test data. For the dataset from Utah, we randomly selected two sets of 35 cases as Center C and D, with a train-test split of 5:1. 
Following the work of~\cite{cms}, to simulate the annotation bias and variation, we applied morphological operations, \textit{i.e.,} open and random erosion to training labels in Center C, and close and random dilation to training labels in Center D. Examples of results are presented in the Supplementary Materials. 
For the local test data, we applied open and close operations, since the aleatoric variation should not occur in the gold standard. 
Besides, we generated a test set of generic data, using 30 cases from Utah with no modification to the gold standard labels. 
All MRIs were resampled to the resolution of 0.6×0.6×1.25 mm and cropped into the size of 256 × 256 centering at the heart region, with Z-score normalization. Random rotation, flip, elastic deformation and Gaussian noise were applied for data augmentation during training. 

\subsection{Implementation}
The segmentation network was implemented using 2D UNet~\cite{Unet}. The structure of prior and posterior encoders was similar to the implementation in~\cite{probU}. The personalized module was built with five convolution layers and SoftPlus activation (please refer to supplementary material for details). Parameters $\alpha,\beta$ and $q$ were set to 0.01, 0.01 and 0.7 (parameter studies are presented in supplementary material), respectively. The networks were trained for 500 epochs with 50 epochs for warm-up. We used the Adam optimizer to update the parameters with a learning rate of 1e-3. The framework was implemented using Pytorch on one Nvidia RTX 3090 GPU.

\subsection{Results}
To validate that the proposed method could simultaneously benefit the global model and personalized models in the scenario of multiple centers and data heterogeneity, we evaluated them in two tasks, \textit{i.e.,} one to test global model on the generic data, and the other to test personalized ones on local data. 
Three global models and three personalized methods were compared, as shown in Table~\ref{tab:comparison}.


\begin{table}[!t]
\caption{\label{tab:comparison}Results (in Dice) and comparisons.
For Global Methods, the global model was used for each center in Task2. 
For Personalized Methods, the results of Task1 were averaged from the four personalized models (for the four centers).
Single: a method that trains four models solely with local data for each center.
Bold text denotes the best results in each column.
}
\small
\centering
		\resizebox{1\textwidth}{!}{
\begin{tabular}{>{\centering\arraybackslash}p{1.8cm}|>{\centering\arraybackslash}p{2cm}|>{\centering\arraybackslash}p{2cm}>{\centering\arraybackslash}p{2cm}>{\centering\arraybackslash}p{2cm}>{\centering\arraybackslash}p{2cm} >{\centering\arraybackslash}p{2cm}}
\hline
\multirow{2}{*}{Method} & \textbf{Task1} & \multicolumn{4}{c}{\textbf{Task2}} \\
\cline{2-7}
&Generic & Center A & Center B & Center C & Center D & mean \\
\hline\hline
 \multicolumn{6}{l}{{Global Methods}} \\
\hline
Swarm~\cite{swarm} &0.843$\pm$0.044 & 0.749$\pm$0.095 & 0.739$\pm$0.091 & 0.839$\pm$0.021 & 0.823$\pm$0.031 & 0.795$\pm$0.054\\
FedProx~\cite{FedProx} &0.869$\pm$0.030 & 0.750$\pm$0.114 & 0.740$\pm$0.086 & 0.851$\pm$0.025 & 0.850$\pm$0.023 & 0.807$\pm$0.056\\
FedCurv~\cite{FedCurv} & 0.866$\pm$0.038 & 0.729$\pm$0.218 & 0.758$\pm$0.093 & 0.837$\pm$0.028 & 0.824$\pm$0.028 & 0.794$\pm$0.081 \\
\hline\hline
 \multicolumn{6}{l}{{Personalized Methods}} \\
\hline
Single & 0.734$\pm$0.143& 0.693$\pm$0.203 & 0.791$\pm$0.054 & 0.708$\pm$0.049 & 0.768$\pm$0.027 & 0.740$\pm$0.076\\
FedRep~\cite{FedRep} & 0.791$\pm$0.054 & \textbf{0.775$\pm$0.218} & 0.764$\pm$0.062 & 0.769$\pm$0.041 & 0.833$\pm$0.029 & 0.788$\pm$0.078  \\
Ditto~\cite{ditto} & 0.781$\pm$0.068 & 0.762$\pm$0.187 & 0.786$\pm$0.062 & 0.763$\pm$0.028 & 0.810$\pm$0.033 & 0.781$\pm$0.070\\
\hline\hline
Ours & \textbf{0.885$\pm$0.027}& 0.772$\pm$0.124 & \textbf{0.793$\pm$0.042} & \textbf{0.874$\pm$0.020} & \textbf{0.873$\pm$0.032} & \textbf{0.836$\pm$0.050}\\
\hline
\end{tabular}}
\end{table}

Our method outperformed all the methods in both of the two tasks. 
Especially, our method improved the Swarm method by $4.2\%$ in Dice, indicating the effectiveness
of the proposed DA network in maintaining the global model from  diverging during local training.
For Task 2, our method not only achieved better results than global methods in all centers, but also set the new state of the art when compared with the personalized methods. 
It is reasonable as the adaptation matrices justify the prediction according to the distribution in each center.
Note that our method achieved comparable mean Dice compared with FedRep and Ditto in Center A and B, but demonstrated significant superiority in Center C and D. 
This was due to the fact that the personalized methods were fine-tuned on each local distribution, thus tended to be affected by aleatoric errors in the annotations which existed in Center C and D.
By contrast, our method was as robust as the global methods for generic test data. 
This advantage was confirmed again by the results in Task 2 for the personalized methods, which were affected by the segmentation bias in Center C and D and were much worse than ours, because our proposed adaptation network was able to learn this bias for local test data.  

\begin{table}[!t]
\caption{\label{tab:ab}Ablation study for the latent representation for the joint data distribution. Results are evaluated in Dice score.}
\small
\centering
		\resizebox{0.9\textwidth}{!}{

\begin{tabular}{>{\centering\arraybackslash}p{1.8cm}|>{\centering\arraybackslash}p{2cm}|>{\centering\arraybackslash}p{2cm}>{\centering\arraybackslash}p{2cm}>{\centering\arraybackslash}p{2cm}>{\centering\arraybackslash}p{2cm}}
\hline
\multirow{2}{*}{Method} & \textbf{Task1} & \multicolumn{4}{c}{\textbf{Task2}} \\
\cline{2-6}
&Generic & Center A & Center B & Center C & Center D \\
\hline
FixedAdapt & 0.864$\pm$0.143 & 0.748$\pm$0.126 & 0.782$\pm$0.067 & 0.808$\pm$0.049 & 0.86$\pm$0.027 \\
ImgAdapt & 0.858$\pm$0.093 & 0.695$\pm$0.188 & 0.735$\pm$0.087 & 0.799$\pm$0.039 & 0.806$\pm$0.045 \\
\hline
Ours & \textbf{0.885$\pm$0.027} & \textbf{0.772$\pm$0.124} & \textbf{0.793$\pm$0.042} & \textbf{0.874$\pm$0.02} &\textbf{0.873$\pm$0.032} \\
\hline
\end{tabular}}
\end{table}

\subsubsection{Ablation study} To validate the importance of the latent representation $z_k$ to the generation of adaptation matrices,
we performed an ablation study, 
by evaluating and comparing the "FixedAdapt" and "ImgAdapt" methods.
In FixedAdapt, the adaptation matrices were fixed during testing (similar to~\cite{gcm}); in ImgAdapt, the adaptation matrices were conditioned on the input image (analogous to~\cite{cms}). 
Table~\ref{tab:ab} presents the results, where our proposed method was evidently better than the other two without information about the local joint distributions. 
This is reasonable as the adaptation network in the two ablated methods tended to overfit on the training data and could not learn the distribution shift in each center. 
With the latent representation $z_k$ that contains distribution information, our method could yield more robust adaptation matrices for test data.

\section{Conclusion}

We have proposed a new method to bridge the gap between global and personalized distributed learning through distribution-conditioned adaptation networks. 
We have validated our method on LA MRI segmentation of multi-center data with both feature and label skew, and showed the method outperformed existing global and personalized methods on both generic data and local data.

%
%
%
\bibliographystyle{splncs04}
\bibliography{refs}

\begin{thebibliography}{10}
\providecommand{\url}[1]{\texttt{#1}}
\providecommand{\urlprefix}{URL }
\providecommand{\doi}[1]{https://doi.org/#1}

\bibitem{bridge}
Chen, H.Y., Chao, W.L.: On bridging generic and personalized federated learning
  for image classification. In: International Conference on Learning
  Representations (2021)

\bibitem{FedRep}
Collins, L., Hassani, H., Mokhtari, A., Shakkottai, S.: Exploiting shared
  representations for personalized federated learning. arXiv preprint
  arXiv:2102.07078  (2021)

\bibitem{MAE}
Ghosh, A., Kumar, H., Sastry, P.: Robust loss functions under label noise for
  deep neural networks. In: Proceedings of the AAAI conference on artificial
  intelligence. vol.~31 (2017)

\bibitem{adapt2}
Goldberger, J., Ben-Reuven, E.: Training deep neural-networks using a noise
  adaptation layer. In: ICLR (2017)

\bibitem{noniid}
Kairouz, P., McMahan, H.B., Avent, B., Bellet, A., Bennis, M., Bhagoji, A.N.,
  Bonawitz, K., Charles, Z., Cormode, G., Cummings, R., et~al.: Advances and
  open problems in federated learning. arXiv preprint arXiv:1912.04977  (2019)

\bibitem{data2}
Karim, R., Housden, R.J., Balasubramaniam, M., Chen, Z., Perry, D., Uddin, A.,
  Al-Beyatti, Y., Palkhi, E., Acheampong, P., Obom, S., et~al.: Evaluation of
  current algorithms for segmentation of scar tissue from late gadolinium
  enhancement cardiovascular magnetic resonance of the left atrium: an
  open-access grand challenge. Journal of Cardiovascular Magnetic Resonance
  \textbf{15}(1),  1--17 (2013)

\bibitem{scaffold}
Karimireddy, S.P., Kale, S., Mohri, M., Reddi, S., Stich, S., Suresh, A.T.:
  Scaffold: Stochastic controlled averaging for federated learning. In:
  International Conference on Machine Learning. pp. 5132--5143. PMLR (2020)

\bibitem{probU}
Kohl, S., Romera-Paredes, B., Meyer, C., De~Fauw, J., Ledsam, J.R., Maier-Hein,
  K., Eslami, S., Jimenez~Rezende, D., Ronneberger, O.: A probabilistic u-net
  for segmentation of ambiguous images. Advances in neural information
  processing systems  \textbf{31} (2018)

\bibitem{LAseg}
Li, L., Zimmer, V.A., Schnabel, J.A., Zhuang, X.: Medical image analysis on
  left atrial lge mri for atrial fibrillation studies: A review. Medical Image
  Analysis p. 102360 (2022)

\bibitem{ditto}
Li, T., Hu, S., Beirami, A., Smith, V.: Ditto: Fair and robust federated
  learning through personalization. In: International Conference on Machine
  Learning. pp. 6357--6368. PMLR (2021)

\bibitem{FedProx}
Li, T., Sahu, A.K., Zaheer, M., Sanjabi, M., Talwalkar, A., Smith, V.:
  Federated optimization in heterogeneous networks. Proceedings of Machine
  Learning and Systems  \textbf{2},  429--450 (2020)

\bibitem{FedBN}
Li, X., Jiang, M., Zhang, X., Kamp, M., Dou, Q.: Fedbn: Federated learning on
  non-iid features via local batch normalization. arXiv preprint
  arXiv:2102.07623  (2021)

\bibitem{blockchain}
Lu, Y., Huang, X., Dai, Y., Maharjan, S., Zhang, Y.: Blockchain and federated
  learning for privacy-preserved data sharing in industrial iot. IEEE
  Transactions on Industrial Informatics  \textbf{16}(6),  4177--4186 (2019)

\bibitem{FedAvg}
McMahan, B., Moore, E., Ramage, D., Hampson, S., y~Arcas, B.A.:
  Communication-efficient learning of deep networks from decentralized data.
  In: Artificial intelligence and statistics. pp. 1273--1282. PMLR (2017)

\bibitem{MAE1}
Natarajan, N., Dhillon, I.S., Ravikumar, P.K., Tewari, A.: Learning with noisy
  labels. Advances in neural information processing systems  \textbf{26} (2013)

\bibitem{Unet}
Ronneberger, O., Fischer, P., Brox, T.: U-net: Convolutional networks for
  biomedical image segmentation. In: International Conference on Medical image
  computing and computer-assisted intervention. pp. 234--241. Springer (2015)

\bibitem{FedCurv}
Shoham, N., Avidor, T., Keren, A., Israel, N., Benditkis, D., Mor-Yosef, L.,
  Zeitak, I.: Overcoming forgetting in federated learning on non-iid data.
  arXiv preprint arXiv:1910.07796  (2019)

\bibitem{cVAE}
Sohn, K., Lee, H., Yan, X.: Learning structured output representation using
  deep conditional generative models. Advances in neural information processing
  systems  \textbf{28} (2015)

\bibitem{adapt1}
Sukhbaatar, S., Bruna, J., Paluri, M., Bourdev, L.D., Fergus, R.: Training
  convolutional networks with noisy labels. arXiv: Computer Vision and Pattern
  Recognition  (2014)

\bibitem{CE}
Szita, I., L{\"o}rincz, A.: Learning tetris using the noisy cross-entropy
  method. Neural computation  \textbf{18}(12),  2936--2941 (2006)

\bibitem{gcm}
Tanno, R., Saeedi, A., Sankaranarayanan, S., Alexander, D.C., Silberman, N.:
  Learning from noisy labels by regularized estimation of annotator confusion.
  In: Proceedings of the IEEE/CVF conference on computer vision and pattern
  recognition. pp. 11244--11253 (2019)

\bibitem{swarm}
Warnat-Herresthal, S., Schultze, H., Shastry, K.L., Manamohan, S., Mukherjee,
  S., Garg, V., Sarveswara, R., H{\"a}ndler, K., Pickkers, P., Aziz, N.A.,
  et~al.: Swarm learning for decentralized and confidential clinical machine
  learning. Nature  \textbf{594}(7862),  265--270 (2021)

\bibitem{data1}
Xiong, Z., Xia, Q., Hu, Z., Huang, N., Bian, C., Zheng, Y., Vesal, S.,
  Ravikumar, N., Maier, A., Yang, X., et~al.: A global benchmark of algorithms
  for segmenting the left atrium from late gadolinium-enhanced cardiac magnetic
  resonance imaging. Medical Image Analysis  \textbf{67},  101832 (2021)

\bibitem{label}
Zhang, H., Valcarcel, A.M., Bakshi, R., Chu, R., Bagnato, F., Shinohara, R.T.,
  Hett, K., Oguz, I.: Multiple sclerosis lesion segmentation with tiramisu and
  2.5 d stacked slices. In: International Conference on Medical Image Computing
  and Computer-Assisted Intervention. pp. 338--346. Springer (2019)

\bibitem{cms}
Zhang, L., Tanno, R., Xu, M.C., Jin, C., Jacob, J., Cicarrelli, O., Barkhof,
  F., Alexander, D.: Disentangling human error from ground truth in
  segmentation of medical images. In: Advances in Neural Information Processing
  Systems. vol.~33, pp. 15750--15762 (2020)

\bibitem{lq}
Zhang, Z., Sabuncu, M.: Generalized cross entropy loss for training deep neural
  networks with noisy labels. Advances in neural information processing systems
   \textbf{31} (2018)

\end{thebibliography}
%




\end{document}